\documentclass[conference]{IEEEtran}
\IEEEoverridecommandlockouts
\usepackage{cite}
\usepackage{amsmath,amssymb,amsfonts}
\usepackage{algorithmic}
\usepackage{graphicx}
\usepackage{textcomp}
\usepackage{xcolor}
\usepackage{fancyhdr}
\usepackage{xcolor}
\usepackage{tikz}
\usepackage{multirow}

\usepackage{siunitx}
\usepackage[a4paper, total={184mm,239mm}]{geometry}
\def\BibTeX{{\rm B\kern-.05em{\sc i\kern-.025em b}\kern-.08em
    T\kern-.1667em\lower.7ex\hbox{E}\kern-.125emX}}

\usepackage{acronym}
\acrodef{EMG}[EMG]{electromyography}
\acrodef{sEMG}[sEMG]{Surface electromyography}
\acrodef{HMI}[HMI]{Human-Machine Interface}
\acrodef{DL}[DL]{Deep Learning}
\acrodef{ML}[ML]{Machine Learning}
\acrodef{SoA}[SoA]{State-of-the-Art}
\acrodef{AP}[AP]{Action Potential}
\acrodef{MUAP}[MUAP]{Motor Unit Action Potential}
\acrodef{MUAPT}[MUAPT]{MUAP Train}
\acrodef{PLI}[PLI]{Power Line Intereference}
\acrodef{MAV}[MAV]{Mean Absolute Value}
\acrodef{WL}[WL]{Waveform Length}
\acrodef{RMS}[RMS]{Root Mean Square}
\acrodef{$k$-NN}[$k$-NN]{$k$-Nearest Neighbors} 
\acrodef{SVM}[SVM]{Support Vector Machine} 
\acrodef{RBF}[RBF]{Radial Basis Function}
\acrodef{RF}[RF]{Random Forest}
\acrodef{HDC}[HDC]{Hyper-Dimensional Computing}
\acrodef{MLP}[MLP]{Multi-Layer Perceptron}
\acrodef{CNN}[CNN]{Convolutional Neural Network}
\acrodef{RNN}[RNN]{Recurrent Neural Network}
\acrodef{LSTM}[LSTM]{Long Short-Term Memory}
\acrodef{TCN}[TCN]{Temporal Convolutional Network}
\acrodef{ReLU}[ReLU]{Rectified Linear Unit}
\acrodef{BN}[BN]{Batch-Normalization}
\acrodef{FC}[FC]{Fully Connected}
\acrodef{MCU}[MCU]{microcontroller unit}
\acrodef{SoC}[SoC]{System on Chip}

\acrodef{MAC}[MAC]{multiply-and-accumulate}
\acrodef{MAE}[MAE]{Mean Absolute Error}
\acrodef{DoF}[DoF]{Degree of Freedom}
\acrodef{DoA}[DoA]{Degree of Actuation}
\acrodef{EMA}[EMA]{Exponential Moving Average}
\acrodef{}[]{}
\acrodef{}[]{}

\newcommand\copyrighttext{%
  \footnotesize \textcopyright 2023 IEEE. Personal use of this material is permitted.  Permission from IEEE must be obtained for all other uses, in any current or future media, including reprinting/republishing this material for advertising or promotional purposes, creating new collective works, for resale or redistribution to servers or lists, or reuse of any copyrighted component of this work in other works.
 
  Published at 2023 Design, Automation \& Test in Europe Conference \& Exhibition (DATE).}

\newcommand{\copyrightnotice}{%
\begin{tikzpicture}[remember picture,overlay,scale=1.00, every node/.style={scale=1.00}]
\node[anchor=south,yshift=10pt] at (current page.south) {\fbox{\parbox{\dimexpr\textwidth-\fboxsep-\fboxrule\relax}{\copyrighttext}}};
\end{tikzpicture}%
}
    
\renewcommand{\baselinestretch}{0.975}
\begin{document}
\bstctlcite{IEEEexample:BSTcontrol}

\title{Energy-efficient Wearable-to-Mobile Offload of ML Inference for PPG-based Heart-Rate Estimation}

\author{\IEEEauthorblockN{Alessio Burrello\IEEEauthorrefmark{1}\IEEEauthorrefmark{2}, Matteo Risso\IEEEauthorrefmark{2}, Noemi Tomasello\IEEEauthorrefmark{2}, Yukai Chen, Luca Benini\IEEEauthorrefmark{1},\\Enrico Macii\IEEEauthorrefmark{2}, Massimo Poncino\IEEEauthorrefmark{2}, Daniele Jahier Pagliari\IEEEauthorrefmark{2}}

\IEEEauthorblockA{
\IEEEauthorrefmark{1} DEI, Università di Bologna, Bologna, Italy 
\IEEEauthorrefmark{2} Politecnico di Torino, Turin, Italy }

\IEEEauthorblockA{Emails: name.surname@unibo.it, name.surname@polito.it}
}

\maketitle

\copyrightnotice

\begin{abstract}
Modern smartwatches often include photoplethysmographic (PPG) sensors to measure heartbeats or blood pressure through complex algorithms that fuse PPG data with other signals. 
In this work, we propose a collaborative inference approach that uses both a smartwatch and a connected smartphone to maximize the performance of heart rate (HR) tracking while also maximizing the smartwatch's battery life. In particular, we first analyze the trade-offs between running on-device HR tracking or offloading the work to the mobile. Then, thanks to an additional step to evaluate the \emph{difficulty} of the upcoming HR prediction, we demonstrate that we can smartly manage the workload between smartwatch and smartphone, maintaining a low mean absolute error (MAE) while reducing energy consumption. 
We benchmark our approach on a custom smartwatch prototype, including the STM32WB55 MCU and Bluetooth Low-Energy (BLE) communication, and a Raspberry Pi3 as a proxy for the smartphone. With our Collaborative Heart Rate Inference System (CHRIS), we obtain a set of Pareto-optimal configurations demonstrating the same MAE as State-of-Art (SoA) algorithms while consuming less energy. For instance, we can achieve approximately the same MAE of TimePPG-Small~\cite{Risso2021} (5.54 BPM MAE vs. 5.60 BPM MAE) while reducing the energy by 2.03$\times$, with a configuration that offloads 80\% of the predictions to the phone. Furthermore, accepting a performance degradation to 7.16 BPM of MAE, we can achieve an energy consumption of 179 uJ per prediction, 3.03$\times$ less than running TimePPG-Small on the smartwatch, and 1.82$\times$ less than streaming all the input data to the phone.
\end{abstract}

\section{Introduction}\label{sec:intro}
\looseness=-1

First-generation wrist-worn devices were equipped mainly with accelerometers to predict daily human activities and assess the user's fitness status.
More recently, thanks to the introduction of new miniaturized sensors and energy-efficient MCUs~\cite{al2018review}, wearable systems started supporting additional personal care functionalities, such as HR monitoring and blood pressure measurement~\cite{nelson2020guidelines}.
Nowadays, the PPG sensors in smartwatches have almost wholly replaced chest bands with Electrocardiograms (ECG) for continuous HR monitoring, being much easier to wear and comfortable for daily usage.
However, PPG sensors, which measure an optical signal related to the blood flow in capillaries, are less accurate and more prone to noise than ECGs. 
The main noise sources are Motion Artifacts (MAs), usually caused by a leakage of light between the skin and the smartwatch during movement.

The problem of reducing the impact of MAs has been extensively addressed in the SoA.
Since the introduction of the first publicly available PPG-based dataset~\cite{troika2014} in 2014, a series of algorithms~\cite{troika2014, spama2016} based on filtering or correlating accelerometer and PPG signals have been proposed to cope with MAs distortion.
These techniques achieve excellent accuracy, but they are often computationally expensive and generalize poorly on subjects not observed during parameters' tuning.

\looseness=-1
Recently, deep learning approaches have demonstrated to i) improve accuracy, ii) improve generalization, and iii) reduce the cost of HR estimation.
In 2019, PPGDalia~\cite{reiss2019deep} introduced the largest publicly available PPG dataset, together with the first deep learning algorithm for HR estimation, based on frequency spectrum extraction and 2D convolutional networks. 
This algorithm obtained superior performance to the SoA but with very high costs for embedded deployment.
Subsequent works on PPG-based HR estimation have given more consideration to deployment aspects but focused only on scenarios in which the prediction is performed entirely on the smartwatch that collects the PPG data~\cite{9583926}, or on a higher-end mobile device~\cite{PPG_survey}.

Conversely, our work considers a more general setting in which both the wearable and the mobile, connected through a BLE link, can perform HR predictions.
To properly map the workload on this multi-device system, we propose CHRIS, a \textbf{C}ollaborative \textbf{H}eart \textbf{R}ate \textbf{I}nference \textbf{S}ystem, which combines and orchestrates classical and DL approaches. 
CHRIS evaluates, for each input sample, which one of the available HR tracking algorithms should be executed, and where (smartwatch or phone), to achieve good accuracy while minimizing energy consumption.
In detail, our contributions are:
\begin{itemize}
    \item We demonstrate that running HR tracking algorithms only on a smartwatch is not always the optimal solution.
    \item We propose CHRIS, a mapping system that uses an Activity Recognition algorithm to differentiate between easy and hard to process PPG windows based on the amount of MAs, consequently selecting an appropriate HR tracking model and dispatching its execution to the smartwatch or to the mobile.
    \item We analyze the results obtained deploying CHRIS on a real system composed by the HWatch~\cite{polonelli2021h}, which features a STM32WB55 as the main MCU, and by a Raspberry Pi3, as a proxy of the phone. We compare CHRIS with single-device solutions in terms of mean absolute error (MAE) and wearable energy. To benchmark the MAE, we use PPGDalia~\cite{reiss2019deep}, the largest publicly available PPG dataset for HR prediction.
\end{itemize}
Thanks to CHRIS, we obtain a rich collection of Pareto points, including solutions that are superior to any single-model/single-device approach.
For instance, we achieve 5.54 Beats Per Minute (BPM) of MAE (roughly equivalent to the 5.60 BPM obtained by a state-of-the-art deep learning model, TimePPG-Small~\cite{Risso2021}), while consuming 22\% less energy compared to an approach that always executes this model on the phone (i.e., streaming all the data to it), and 44.3\% less compared to always executing it on the smartwatch. This result is obtained by executing a deep network (on the phone) only for particularly difficult inputs, and replacing it with a much simpler deterministic model, executed on the smartwatch, for easy samples.

\section{Background \& Related Work}
\label{sec:background_related}
\subsection{Photopletismographic signal}
The photoplethysmographic signal measures the light absorption variations of blood vessels~\cite{tamura2014wearable}.
A PPG sensor measures 1 to 3 components, acquired through Light-Emitting Diodes (LEDs) that continuously emit light to the skin at different frequencies (green, red, and infrared), and a corresponding photodetector that measures the amount of the reflected light caused by the blood flow, which follows the HR periodicity.
In particular, a larger blood volume implies a more significant attenuation of the light emitted by the LED. 
Therefore, the photodetector will generate a lower current that can be associated with a heartbeat.
As anticipated in Sec.~\ref{sec:intro}, the simplicity of this sensor comes with the significant challenge of managing the so-called MAs generated by light leaking between the skin and the device due to the subject's movement.
The standard approach to tackle the MAs is to leverage sensor fusion, using additional inputs from an inertial sensor (accelerometer), whose correlation with PPG is used to mitigate the MAs. 
More details can be found in~\cite{PPG_survey}.

\subsection{Collaborative Inference}
\label{subsec:collaborative_learning}
Collaborative inference refers to the execution of ML inference tasks on a distributed system composed of multiple edge and cloud devices~\cite{Kang2017}.
It is a specialization of the more general edge-cloud offloading paradigm~\cite{Cuervo2010}, and was initially proposed to split the execution of large deep neural networks (e.g., layer-wise or input-wise) in a way that optimizes the total response time of the system, or the energy consumption of the edge node~\cite{Kang2017,JahierPagliari2020b}. 
More recently, collaborative inference has been extended to scenarios in which the involved devices can execute heterogeneous ML models based on their computational capabilities~\cite{Shlezinger2021}. 
While this approach can generally involve any number of connected devices and various types of relationships among them, the most common embodiment includes one low-power, low-performance ``in-field'' device offloading computational tasks to a more powerful ``remote device''~\cite{Kang2017}. In our work, these two roles are taken by the PPG-equipped wearable and by the mobile, respectively. 

\subsection{The HWatch Platform}
\label{subsec:HWatch}
Although our proposed method, CHRIS, is orthogonal to the underlying hardware platform, in this work, we benchmark it on the HWatch~\cite{polonelli2021h}, a custom board specifically designed to develop wrist-worn applications with the form factor of a smartwatch. 
The watch includes a STM32WB55RGV6 System-on-Chip (SoC) from ST Microelectronics~\footnote{https://www.st.com/en/microcontrollers-microprocessors/stm32wb55rg.html}, named STM32WB hereafter. 
The STM32WB is composed of two fully independent cores, an Arm\textsuperscript{\textregistered} Cortex\textsuperscript{\textregistered}‐M4 core running at 64 MHz (application processor) and an Arm\textsuperscript{\textregistered} Cortex\textsuperscript{\textregistered}‐M0+ core at 32 MHz (network processor).
The STM32WB also includes a Radio-Frequency (RF) transceiver with a radio stack compliant with BLE 5.0 standard. 
The power supply sub-system of the board exploits a TPS63031 buck-boost converter from Texas Instruments. 
This converter reaches 90\% efficiency during sensor acquisition and processing modes.
A Li-Ion 370~mA@3.7V battery is used by the TPS63031 as primary power source.

Besides MCU and power supply, two other sensors, a MAX30101~\footnote{https://www.maximintegrated.com/en/products/interface/sensor-interface/MAX30101.html}, and an LSM6DSM~\footnote{https://www.st.com/resource/en/datasheet/lsm6dsm.pdf}, are included in the system. 
The former is a low-power pulse oximeter and HR monitor (i.e., PPG) module, while the latter is a 6-D inertial measurement unit with a ML core optimized for executing Random Forests (RFs). 
They are connected with the MCU using respectively I2C and SPI digital busses.
The hardware power consumption of the different components is analyzed in~\cite{polonelli2021h}.

\subsection{Related Work}
\label{subsec:related}
The seminal work that paved the way for PPG-based HR monitoring was published in 2014~\cite{troika2014}, together with the first open-access dataset (the 12-subject SPC).
The authors proposed an algorithm based on signal decomposition, spectrum extraction, and peak tracking (TROIKA), achieving 2.34 BPM of MAE.
In successive years, many works have been published which use similar processing chains to improve the results on the SPC dataset, reducing the MAE down to 0.89 BPM~\cite{spama2016}. 

Despite the impressive results, these processing pipelines hardly generalize to new and unseen data~\cite{reiss2019deep}.
For this reason, in 2019, the authors of~\cite{reiss2019deep} introduced the first deep learning approach to this task, a 2D-CNN with a pre-processing step to extract the signal's spectrum.
In \cite{reiss2019deep}, the authors also introduced PPGDalia, the largest PPG dataset publicly available, which is also used in this work.
In the following few years, many different deep learning based solutions have been published~\cite{cornet2019, shyam2019ppgnet,chung2020deep, rocha2020binary}.
However, only a very limited number of works coupled the powerful generalization of DNNs with studying their deployment on real devices.
In particular, \cite{10.1145/3487910, 9583926} introduced two collections of temporal convolutional networks, QPPG and TimePPG, with different accuracy and complexity, which could be deployed on low-power MCUs.
Furthermore, \cite{10.1145/3487910} also introduced an arbitration criterion to combine multiple DNNs at runtime.
These works, however, only considered the scenario in which the HR tracking task is \textit{entirely} handled by the smartwatch.

In our work, we extend the concept of an ensemble of classifiers of~\cite{10.1145/3487910} with collaborative inference, allowing the usage of different models on board and the possibility to offload the workload to a more powerful connected device, opening more opportunities for energy minimization.
To the best of our knowledge, we are the first to investigate collaborative inference for this task.
\section{Collaborative Heart Rate Inference System}
\label{sec:chris}
%
CHRIS is a simple runtime executing on the smartwatch, which optimizes the HR tracking error and energy consumption based on: i) a proxy of the ``difficulty'' of the current input, ii) the connection status, and iii) a user-defined error or energy constraint.
To achieve this goal, CHRIS embodies multiple models for HR estimation: for each new estimation, a model is chosen from the set and either executed locally on the wearable or offloaded to a connected phone.

\begin{figure}[t]
  \centering
\includegraphics[width=1.0\linewidth]{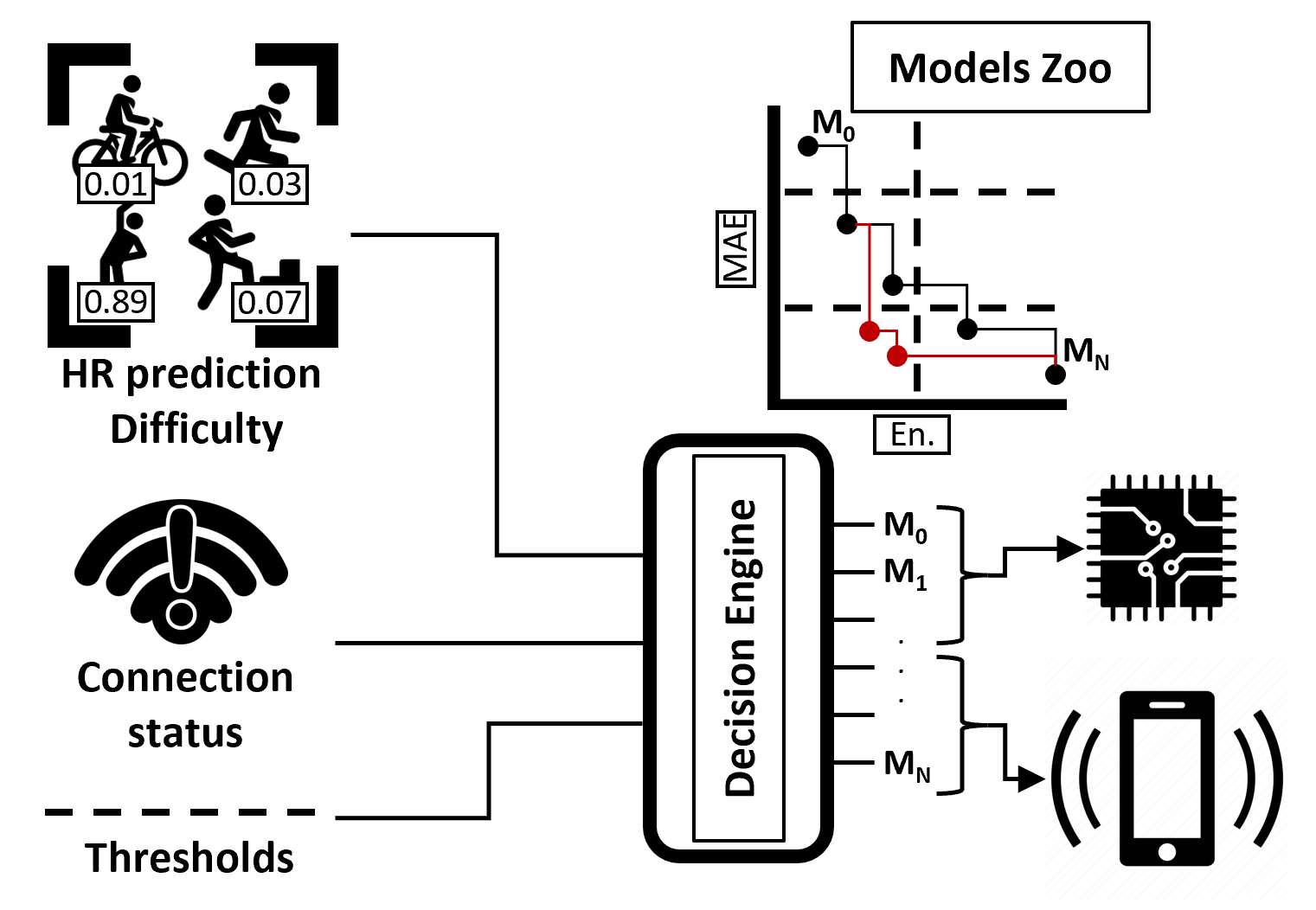}
  \caption{The CHRIS framework. The main components are the Decision Engine and the Models Zoo. The framework outputs are the model to be executed and the platform to be used (phone or smartwatch).}
  \label{fig:CHRIS}
  \vspace{-0.1cm}
\end{figure}

Fig.~\ref{fig:CHRIS} depicts the system, its inputs, and the network dispatching.
The main components are the Decision Engine, which is utilized to choose which model to execute and on which device (Fig.~\ref{fig:CHRIS_arbiter}), and a Models Zoo, a collection of models each characterized by its error and by the on-board and on-phone energy consumption (Table~\ref{tab:ensemble}).

\subsection{CHRIS Configurations}\label{sec:models}
\begin{table}[t]
  \centering
  \caption{Example of models used to construct CHRIS configurations.}
\begin{tabular}{lllll}
\hline
              &           & \multicolumn{3}{c}{Energy [mJ]} \\ 
              & MAE [BPM] & Board     & Phone     & BLE     \\ \hline \hline
AT            & 10.84     & 0.23      & 1.61      & \multirow{3}*{0.52}    \\ 
TimePPG-Small & 5.63      & 0.543     & 5.54      &     \\
TimePPG-Big   & 4.88      & 41.11     & 25.60     &     \\ \hline
\end{tabular}
  \label{tab:ensemble}
  \vspace{-0.5cm}
\end{table}
\begin{table}[t]
\centering
  \caption{Configurations stored inside CHRIS.}
\begin{tabular}{llllll}
\hline
      & MAE [BPM] & E. [mJ] & Models             & Diff. & Exec. \\ \hline
$C_1$ & 10.11          &     0.92       & [AT, TimePPGSmall] & 9           & Local     \\
$C_2$ &      10.05     &        0.87    & [AT, TimePPGBig]   & 9           & Hybrid    \\
…     &           &            &                    &             &           \\
$C_N$ &     5.11      &     40.05       & [AT, TimePPGBig]   & 1           & Local     \\ \hline
\end{tabular}
  \label{tab:conf}
  \vspace{-0.5cm}
\end{table}
The first component of CHRIS is a collection of operating \textit{configurations} that are profiled offline, storing the results of such profiling on the smartwatch.
We call configuration a group of 2 HR prediction models, where only one model is executed for each input window.
Each configuration is characterized in terms of average energy consumption, average MAE, difficulty threshold used by the decision engine (detailed below), and type of execution (entirely on the smartwatch or hybrid).
Average energy and MAE are estimated on a profiling dataset.
Each configuration includes a more accurate but more energy-hungry model and a less accurate but more efficient one.

Table~\ref{tab:ensemble} reports the characterization of individual models, whereas Table~\ref{tab:conf} shows an example of the configurations' profiling information stored in the smartwatch MCU memory.

\looseness=-1
Given a configuration and an incoming input window, the \emph{difficulty threshold} is used to decide which model to employ for estimating the HR. 
We consider 9 different difficulty levels, corresponding to the activities performed by the subjects in the PPGDalia dataset~\cite{reiss2019deep}. These activities includes i) sitting, ii) ascending/descending stairs, iii) playing table soccer, iv) cycling, v) driving a car, vi) having lunch, vii) walking, viii) working and ix) resting.
As discussed in~\cite{10.1145/3487910}, each activity can be associated with a different quantity of movement, which corresponds to a different amount of MAs in the PPG signal, and, therefore, to a different HR estimation difficulty.
For instance, playing table soccer is associated with lots of MAs, due to the sudden movements of the arms. We can therefore order the activities by difficulty based on the average accelerometer signal energy, as proposed in~\cite{10.1145/3487910}, associating them with a cardinal number from 1 to 9, where lower numbers correspond to a lower degree of MAs.
The difficulty threshold in CHRIS then determines the \textit{maximum difficult level for which the simpler and more efficient model of a given configuration will be employed for HR tracking}. For instance, if the threshold is set to 4 for a given configuration, then CHRIS will invoke the smaller model when it predicts that the subject is executing one of the 4 ``simplest'' activities.
The more complex model will be used for inputs that correspond to the remaining five activities.

To reduce the overhead of identifying the appropriate model to run and the target device, we store configurations ordered by energy and MAE.
%
%
In this way, a single linear-complexity pass-on is sufficient to retrieve the optimal configuration given the user-defined constraints, which are discussed below.

\subsection{CHRIS Decision Engine}
\label{sec:decision}
\subsubsection{Constraints-dependent configuration selection}
\begin{figure}[t]
  \centering
\includegraphics[width=1.0\linewidth]{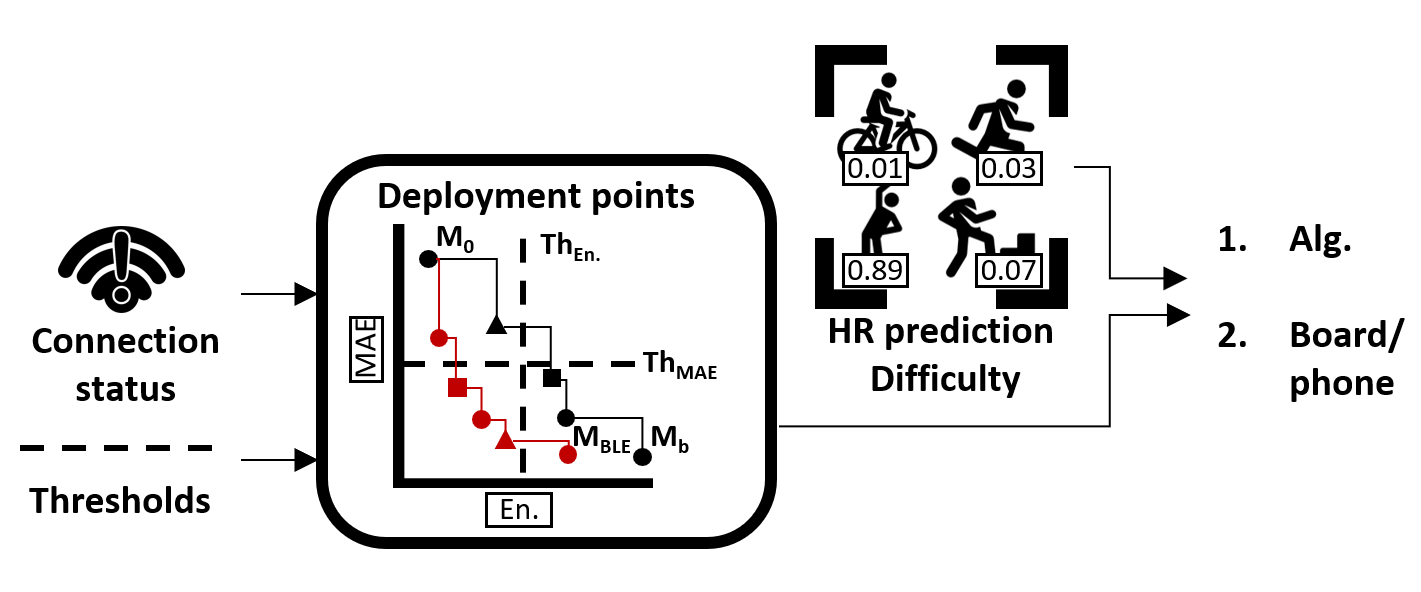}
  \caption{Decision Engine of CHRIS. Using as input the connection status and a user-defined threshold, it selects a configuration and decides which model to run and where based on the current input. }
  \label{fig:CHRIS_arbiter}
  \vspace{-0.5cm}
\end{figure}

Fig.~\ref{fig:CHRIS_arbiter} shows the internal structure of the decision engine of CHRIS. 
The connection status and a user-defined threshold are used to choose a configuration from the profiling table stored in the MCU memory.
The connection status is used to limit the search to feasible configurations, i.e., only the \textit{local} ones (in which both models run on the smartwatch) when the BLE link is not available, filtering out \textit{hybrid} configuration (in which the largest model runs on the phone).
The user-defined threshold, instead, identifies a specific configuration within the feasible set. It can be defined either as a maximum expected MAE ($Th_{\text{MAE}}$, horizontal line in Fig~\ref{fig:CHRIS_arbiter}), in which case CHRIS identifies the configuration with the lowest energy consumption achieving a MAE smaller than $Th_{\text{MAE}}$, or as a maximum expected energy consumption ($Th_{\text{En.}}$, vertical line in Fig~\ref{fig:CHRIS_arbiter}), in which case CHRIS selects the configuration with the best MAE among those consuming less than $Th_{\text{En.}}$.
These user-defined thresholds represent a soft constraint that will be respected only when the on-field input data is distributed similarly to the dataset used for profiling the configurations.

\subsubsection{Input-dependent model selection}
Having selected a configuration (i.e. a pair of HR tracking models) as discussed above, CHRIS works by automatically assigning each input window to one of the two models based on the estimated amount of motion artifacts.
Accordingly, the decision is performed based on accelerometer data through a simple activity recognition model that assigns each input window to one of the 9 activities described in Sec. \ref{sec:models}. Any classification model can be used for this step, but in this work we select a small Random Forest (RF) which, besides reaching a good accuracy, allows us to deploy the activity recognition classifier directly on the specialized ML core of the LSM6DSM accelerometer, relieving the main MCU from this task and saving power accordingly.
As output, the RF provides the activity performed by the subject, which is compared with the decision threshold discussed in Sec.~\ref{sec:decision}.
If the predicted activity has a lower ID than the threshold, the input window is fed to the simplest model of the pair.
Otherwise, the most complex model is executed.
In all our experiments, we take into account the impact of RF mispredictions when computing the actual MAE and energy of a configuration. 
Nevertheless, since the RF consistently achieves an accuracy greater than 90\% in discerning easy from difficult activities, we found that occasional mispredictions do not affect the overall functionality of CHRIS significantly.

\subsection{Benchmark HR tracking and activity recognition models}\label{sec:hr_models}

In this work, we consider three individual HR predictors (a simple classical algorithm and two deep learning models) as the basis for building CHRIS configurations. This yields a total of 60 possible configurations (10 difficulty levels, three combinations of 2 out of the 3 models, and two possible destinations for the execution of the most complex model, i.e.,  wearable or mobile).
Out of these, 30 are Pareto-optimal, and their profiling results are stored in the MCU. Notice that, despite the large number of configurations, the smartwatch only needs to store (at most) 3 HR tracking models in its local memory. Hence the memory overhead of CHRIS is limited. Moreover, we underline that the three models considered in this work are just examples since, in general, CHRIS's functionality is orthogonal to the characteristics of the individual HR predictors.

More specifically, our simplest HR predictor is based on the Adaptive Threshold (AT) method described in~\cite{RollingMean}. It computes the rolling mean of the signal over a window of 24 samples. Then, \textit{regions of interest}, i.e., regions where the raw signal is higher than the mean, are identified. The largest values of each region are identified as peaks of the signal. Lastly, the distance between two successive peaks is associated with the HR. AT requires only $\approx$ 3k operations per input window, but it is also the least accurate model, achieving 10.99 BPM of MAE on the PPGDalia dataset.

The two deep learning models are called TimePPG-Small and TimePPG-Big, and are taken from~\cite{10.1145/3487910}.
They are two Temporal Convolutional Networks (TCNs), i.e., 1D convolutional neural networks with \emph{dilation}, an additional hyper-parameter that allows increasing the receptive field without increasing the network complexity by inserting a fixed gap between the input activations convolved with the weights' filters. 
Both models have a modular structure with 3 blocks, composed of three convolutional layers each, two with dilation $>$ 1, and one with stride = 2 (for a total of 9 convolutional layers in each network). The difference between TimePPG-Small and TimePPG-Big is in the number of filters of each layer, which has been optimized by means of a NAS algorithm. Accordingly, TimePPG-Small has 5.09k parameters and performs 77.63k operations per prediction, achieving a MAE of 5.6 BPM on Dalia. TimePPG-Big, instead, is characterized by 232.6k parameters and 12.27M operations, which allow it to obtain a lower MAE of just 4.87 BPM on the same dataset.
Note that while the MAEs of the two networks are considerably lower than AT, their complexity in terms of number of operations is 25.9$\times$ and 4090$\times$ higher, respectively.
Overall, the three considered HR tracking algorithms span over 6 BPM of MAE, and three orders of magnitude in complexity, providing a wide optimization space to CHRIS.

Lastly, the RF used as an activity recognition model is made of 8 trees with a maximum depth of 5.
It is fed with a set of 4 features extracted from the three accelerometer axes and selected by performing a grid search over common statistical features. The 4 selected predictors are: mean, energy, standard deviation, and number of peaks (i.e., number of discrete derivative sign changes).

\section{Experimental Results}
\label{sec:results}

\subsubsection{Platforms}
\label{sec:platforms}
We use the HWatch as a target embedded system for our experiments (Sec. \ref{subsec:HWatch}) and a Raspberry Pi3 equipped with an Arm\textsuperscript{\textregistered} Cortex\textsuperscript{\textregistered}-A53 core as a proxy of the typical processors found in smartphones.
CHRIS and the prediction models are deployed on the STM32WB55 present in the HWatch using the STM neural network deployment toolchain X-CUBE-AI~\cite{xcubeai}.
On the Raspberry Pi3, deep learning models are executed with the TensorFlow Lite interpreter.
TimePPG-Small and -Big are quantized to 8bit using quantization-aware training before deployment for both target platforms.

\subsubsection{Dataset}
\label{sec:dataset}
To train the HR-prediction models and the difficulty detector, we employ data from PPGDalia \cite{reiss2019deep}. 
The dataset comprises 37.5 hours of data recorded from 15 subjects.
Each subject performs eight (+ rest) different daily activities, during which the PPG data is recorded together with 3D-accelerometer data and the golden HR values collected through an ECG chest band.
We employ the acceleration data and the activity labels to train the difficulty detector, whereas we use PPG data, acceleration data, and HR golden values to train the HR tracking neural networks (TimePPG-Big and TimePPG-Small).
We split the data into 5 folds, each composed of three subjects: in each iteration, we use 4 folds for training, two subjects of the last fold as validation, and the last subject for testing. We then rotate each subject within the latter fold as test, before switching to the next iteration.
The raw acceleration and PPG data are sampled at 32 Hz. We separate the original time series into windows of 256 samples (8 seconds) with a stride of 64 (2 seconds) before feeding them to the models.
All experiments are performed using Python 3.6 and PyTorch 1.6.

\subsection{Individual Models Deployment}
\begin{figure}[t]
  \centering
\includegraphics[width=1.0\linewidth]{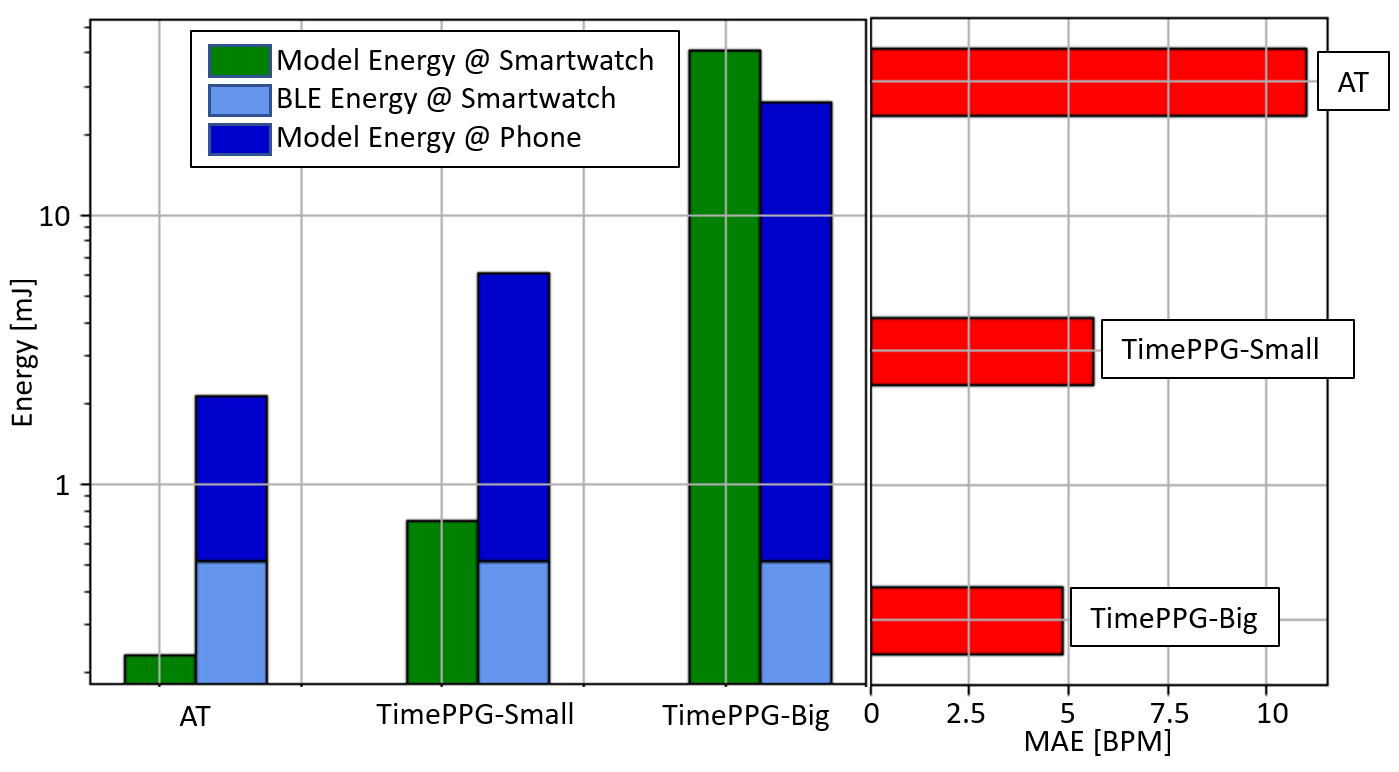}
  \caption{Baseline models. On the left is the energy consumption of the three models. On the right is the average MAE over PPGDalia dataset.}
  \label{fig:baselines}
  \vspace{-0.5cm}
\end{figure}
\begin{table*}[ht]
  \centering
  \caption{Deployment of baseline models on the STM32WB55 and on the Raspberry Pi3.}
\begin{tabular}{l|lll|ll|l}
\hline 
&                   \multicolumn{3}{c|}{STM32WB55} & \multicolumn{2}{c|}{Raspberry Pi3} & \\
 &                   Cycles  & Time [ms] & Energy [mJ] & Time [ms] &Energy [mJ] & MAE [BPM] \\\hline\hline
AT      &  100k        & 1.563          & 0.234 & 1.00 & 1.60 &10.99   \\
TimePPG-Small       & 1.365M    & 21.326    &   0.735    & 3.45 & 5.54 & 5.60      \\
TimePPG-Big   & 103.16M   & 1611.88   &  41.11    & 15.96 & 25.60 & 4.87      \\
Bluetooth       & n.a.       & 10.240     & 0.52   & n.a.& n.a.& n.a.    \\\hline 
\multicolumn{6}{l}{ STM32WB55 MCU Frequency = 64 MHz. Raspberry Pi3 Frequency = 600 MHz.}\\
\end{tabular}
  \label{tab:results}
  \vspace{-0.5cm}
\end{table*}
Fig.~\ref{fig:baselines} reports the energy consumption and MAE of each model. Specifically, we report the total computation energy on the HWatch MCU (green bar, including also the energy spent in idle between two subsequent predictions), the computation energy on the phone (dark blue), and the BLE transmission energy (light blue). Note that the latter is fixed since the input samples dimension does not depend on the selected HR tracking model. Numerical results are also reported in Table~\ref{tab:results}, together with the execution time of each model on both platforms and the number of clock cycles on the HWatch.

As expected, AT is not only the least accurate algorithm but also the most efficient one to execute on the HWatch (0.234 mJ vs. 0.735 mJ and 41.11 mJ of the two DNNs).
Offloading the execution of this algorithm to a smartphone is thus clearly sub-optimal, since both the energy consumed by the watch (for streaming data through BLE) and by the Cortex-A core would be higher (0.234 mJ vs. 0.519 mJ and 1.604 mJ respectively).

TimePPG-Small achieves a good compromise between MAE and energy.
The consumption of this algorithm on the HWatch is 3.1$\times$ higher than the one of AT, for a 1.96x reduction in MAE (5.6 BPM vs 10.99 BPM).
Therefore, running HR tracking on the smartwatch with this model is convenient from the perspective of the whole system's energy. On the other hand, if the objective is to optimize the energy consumed by the smartwatch (often the most critical one from the point of view of the total lifetime of the system), offloading the HR tracking to a phone is slightly more convenient (0.735 mJ for execution vs. 0.519 mJ for BLE transmission).

Lastly, TimePPG-Big is the most accurate and energy-hungry model; it consumes 41.07 mJ per prediction on the HWatch, a value more than 2 orders of magnitudes higher than AT, in order to reach a 2.25x lower MAE (4.87 BPM).
For this algorithm, local execution on the smartwatch is always sub-optimal, both in terms of total system energy and considering only the smartwatch consumption, since BLE transmission requires the usual 0.519 mJ, while running the model on the Cortex-A consumes 25.60 mJ.
Overall, these results demonstrate that the choice between local and offloaded execution is not trivial and depends on the target model, which in turn should be selected based on the desired maximum HR tracking error or consumption.

\subsection{CHRIS Exploration of MAE vs. Energy}
\begin{figure}[t]
  \centering
\includegraphics[width=1.0\linewidth]{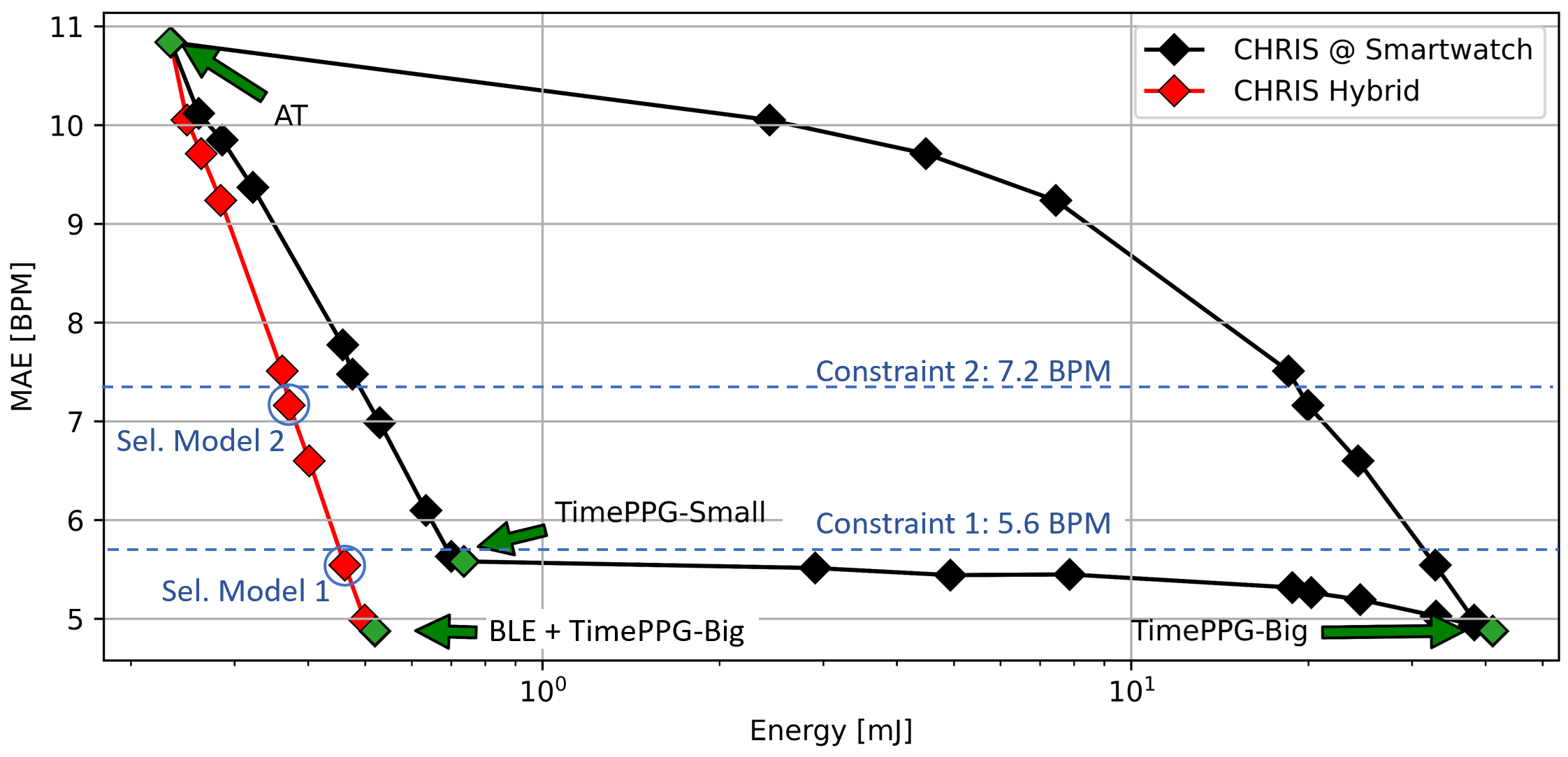}
  \caption{CHRIS' configurations in the space MAE vs. Energy. In red are the points that involve the execution on the smartphone.}
  \label{fig:CHRIS_results}
\end{figure}
\begin{figure}[t]
  \centering
\includegraphics[width=1.0\linewidth]{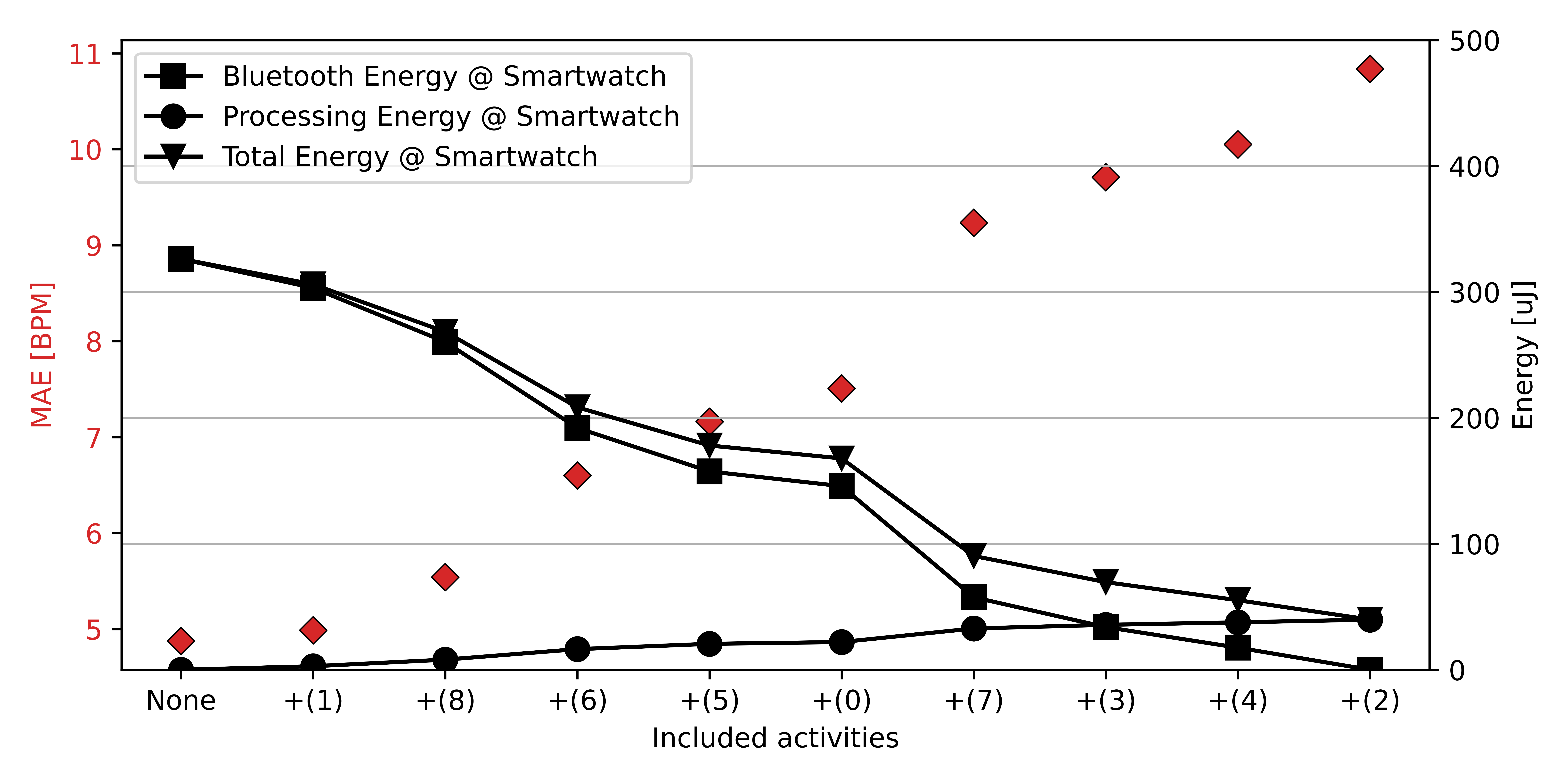}
  \caption{Energy and MAE while varying the number of "easy" and "difficult" human activities.}
  \label{fig:CHRIS_zoom}
  \vspace{-0.5cm}
\end{figure}
Fig.~\ref{fig:CHRIS_results} shows all the possible solutions covered by CHRIS in the space MAE vs. smartwatch energy. We focus on the smartwatch consumption because it is where CHRIS can make a bigger difference; mobiles execute tens of tasks simultaneously, thus HR tracking has a smaller impact on the overall battery life.
The green diamonds in the figure correspond to the baseline solutions described above, and in particular, the \emph{BLE + TimePPG-Big} point corresponds to a solution that always offloads HR tracking to the mobile; since in this experiment we do not optimize the smartphone energy, offloaded samples are always processed with the most accurate model available (TimePPG-Big), which yields the lowest possible MAE.

Red and black diamonds connecting individual models correspond to the intermediate solutions provided by CHRIS, combining two HR tracking algorithms. Each point corresponds to a different ``difficulty threshold''.
In this experiment, the hybrid solution, which combines the local execution of AT and the remote execution of TimePPG-Big, Pareto-dominates the others (red points). Given a user-defined MAE (energy) threshold, CHRIS will thus select the topmost (rightmost) red point within the half-space defined by the horizontal (vertical) line corresponding to the constraint.
For example, given an MAE constraint of 5.60 BPM, (the same value obtained by TimePPG-Small alone) shown as \textit{Constraint 1} in the figure, CHRIS will select \textit{Sel. Model 1}. This corresponds to a combination of AT and TimePPG-Big with a difficulty threshold of 8, in which $\sim$ 80\% of the windows are offloaded to the phone. This combination yields a MAE of 5.54 BPM, while reducing the energy consumption on the smartwatch by 2.03$\times$ compared to running TimePPG-Small locally. 
Relaxing the MAE constraint to 7.2 BPM (\textit{Constaint 2}),  instead, CHRIS will select a configuration with a difficulty threshold of 6 (\textit{Sel. Model 2}), further reducing the average energy consumption to 179 uJ per prediction, 3.03× less than running TimePPG-Small on the smartwatch, and 1.82× less than streaming all the input data to the phone.
Lastly, if the BLE connection to the smartphone is lost (excluding the red points from the feasible solutions), CHRIS would still find 19 Pareto points spanning from 4.87 BPM to 10.99 BPM of MAE and from 0.234 mJ to 41.07 mJ of energy consumption, by combining AT with TimePPG-Small or TimePPG-Small with TimePPG-Big.

Fig.~\ref{fig:CHRIS_zoom} shows in detail how the MAE and the energy consumption breakdown change for the red curve of Fig.~\ref{fig:CHRIS_results} when adding more activities to the ``easy'' group, i.e., processing more windows on the smartphone.
While the variation of energy and MAE is almost linear in this graph, we underline that these results are obtained from the PPGDalia dataset, where all the activities are identically represented. We have, in other words, an equal number of inputs in which the subject is running compared to the ones in which they are sitting or working.
In a real-world scenario, the static and easier-to-process inputs would likely be many more compared to the ones with high movement and, therefore, CHRIS would achieve even better results, resorting to offloading only in rare cases.
\section{Conclusions}
\label{sec:conclusions}
An efficient and accurate prediction of a subject's HR using wearables is key for many personal care applications.
In this work, we introduced CHRIS, a new inference system that exploits the synergy between two connected platforms, a smartwatch and a smartphone, to explore the trade-off between energy consumption and HR tracking error.
Based on the connection status, a user-defined error and/or energy constraint, and an estimate of the input difficulty, CHRIS combines two HR prediction algorithms, executed on the smartwatch or on the phone, achieving up to 2.03x energy reduction on the former with respect to a single-model \& single-device solution, with no impact on the tracking accuracy.

\tiny
\bibliographystyle{IEEEtran}
\bibliography{bstctl,references}

\end{document}